\documentstyle[aps,prl]{revtex}
\begin{document}
\input epsf.sty

\twocolumn[\hsize\textwidth\columnwidth\hsize\csname %
@twocolumnfalse\endcsname

\title{Isotropic Spin Wave Theory of Short-Range Magnetic Order}

\author{Alexander Sokol$^1$}
\address{Department of Physics, University of Illinois at
Urbana-Champaign, Urbana, IL 61801}
\author{Rajiv R.P. Singh}
\address{Department of Physics, University of California, Davis,
CA 95616}
\author{Norbert Elstner}
\address{Service de Physique Th\'{e}orique, CEA-Saclay, 91191
Gif-sur-Yvette Cedex, France}

\date{February 7, 1996}

\maketitle

\begin{abstract}
We present an isotropic spin wave (ISW) theory of short-range order in
Heisenberg magnets,
and apply it to square lattice $S\!=\!1/2$ and $S\!=\!1$
antiferromagnets. Our theory has three identical (isotropic)
spin wave modes,
whereas the conventional spin wave theory has
two transverse and one longitudinal mode.
We calculate temperature dependences
of various thermodynamic
observables analytically and find good (several per cent)
agreement with independently obtained numerical results in a broad
temperature range.
\end{abstract}
\phantom{.}

\bigskip

]

\narrowtext
\pagebreak

{\bf Introduction}.
We study short range magnetic order
in Heisenberg antiferromagnets described by the following
Hamiltonian:
\begin{equation}
{\rm \hat H} = J \sum_{\langle {\bf rr'} \rangle}
{\bf S_r} {\bf S}_{\bf r'},
\label{H}
\end{equation}
where $\langle {\bf rr'}\rangle$ denotes nearest neighbor pairs.
The critical behavior of square lattice Heisenberg antiferromagnets
is known: they have staggered magnetic order
at $T=0$, and exponentially
divergent correlation length in the limit $T\to0$. The purpose of our
paper is not to add anything to understanding their asymptotic properties
for $T\to0$, but rather to
develop an accurate analytical theory for calculating observables
at nonzero temperature, where the correlation length is finite and the
system has short range, but not long range, order.
Recent advances in numerical methods for
quantum magnets provided us with detailed numerical data to test our
analytical predictions.

Our theory is applicable in one and two dimensions.
In this paper, we focus on $S\!=\!1/2$ and $S\!=\!1$
square lattice antiferromagnets, such as
$\rm La_2CuO_4$ and $\rm Sr_2CuO_2Cl_2$
(S=1/2) and $\rm La_2NiO_4$ (S=1).

We obtained numerical results for using high temperature
series expansions for the following quantities:
(i) static susceptibility $\chi({\bf q})=
{\bf M}({\bf q})/{\bf H}({\bf q})$
for arbitrary ${\bf q}$, (ii)
equal time spin correlator $S({\bf q})=\langle {\bf S_q S_{-q}}\rangle$
for arbitrary ${\bf q}$, (iii) correlation length  $\xi(T)$,
(iv) uniform susceptibility $\chi_0(T)$, and
(v) internal energy $U(T)$.
The latter three quantities
are derived from the former two, and
agree with the earlier Monte Carlo
\cite{Ding:Makivic,Sandvik:Scalapino,Greven:mc} and series expansions
\cite{EGSS} data,
and with the experiment \cite{Keimer:lco,Greven:scoc,Greven:S=1}.
Since it is unlikely for different experimental and numerical
methods to have identical systematic errors,
we believe that the numerical results are
accurate and provide a reliable test of the analytical theory.
\newpage

{\bf Isotropic versus Conventional Spin-Wave Theory}.
Magnetic excitations of the Heisenberg model are interacting spin
waves. Their interaction (mode coupling) is essential not
only for dissipation, but also for dynamically
generating one longitudinal and two transverse modes in the
low-temperature limit, as required by the symmetry of the ordered
phase.
Mode coupling is accurately captured by the
quantum nonlinear sigma (QNL$\sigma$) model,
but only at the expense of limiting
its applicability to the temperature range where $\xi\gg 1$
($\xi$ is the correlation length). Here and in what follows we assume
the units where lattice spacing $a=1$ and exchange constant $J=1$.

Recent detailed calculations \cite{CSY}
show that for $T>\rho_s$ ($\rho_s \sim JS^2$ is spin stiffness),
where most of the experimental and numerical data exists,
the effect of mode coupling is rather weak. This observation
led us to explore approximations which ignore mode
coupling, in which case the theory
no longer requires that $\xi\gg 1$.

{\bf Spin Wave Spectrum}.
We linearize the equations of motion written in terms of variables
defined on nearest neighbor bonds:
\begin{eqnarray}
{\bf m}_{\bf rr'}&=&{\bf S_r}+{\bf S_{r'}},\nonumber \\
{\bf n}_{\bf rr'}&=&{\bf S_r}-{\bf S_{r'}},\label{vars}\\
{\bf l}_{\bf rr'}&=&{\bf S_r}\times{\bf S_{r'}},\nonumber
\end{eqnarray}
which form an orthogonal triad for each bond, and seek canonical
variables as linear combinations of $({\bf n_{rr'}},{\bf l_{rr'}})$.

According to the exact solution of the classical Heisenberg chain by
Fisher \cite{Fisher}, the thermodynamic properties of a chain
are identical to those of an isolated pair of classical spins.
Because $({\bf n_{rr'}},{\bf l_{rr'}})$ are exact canonical variables for
an isolated classical spin pair,
a linear theory in terms of these variables must
be able to reproduce exactly the thermodynamics of a classical chain
at arbitrary temperature.
Furthermore, for a system which is sufficiently close to ferro- or
antiferromagnetic long range order (i.e. $\xi\gg 1$),
the variable set (\ref{vars})
is suitable for the derivation of the corresponding ferro- or
antiferromagnetic quantum nonlinear sigma model, which makes the
correspondence between the sigma model theory and our isotropic spin wave
theory more transparent.

The equations of motion written for ${\bf n_{rr'}}$ and
${\bf l_{rr'}}$ on nearest neighbor bonds
\begin{eqnarray}
\frac{\partial \bf n_{rr'}}{\partial t} &=&
- \sum_{\delta} \left(
{\bf l_{r,r+\delta}} - {\bf l_{r',r'+\delta}} \right) \nonumber\\
\frac{\partial  \bf l_{rr'}}{\partial t} &=&
\sum_{\delta} \left(
C_{\bf r+\delta,r'} {\bf S_r} -
C_{\bf rr'} {\bf S_{r+\delta}} - \right. \label{motion} \\
&& \left. {\bf S_{r'}} C_{\bf r'+\delta,r} +
{\bf S_{r'+\delta}} C_{\bf r'r}
\right), \nonumber
\end{eqnarray}
where $C_{\bf rr'} = {\bf S_r S_{r'}}$, are linearized by
replacing
\begin{equation}
\sum_{\delta} C_{\bf r+\delta,r'} \to Z C_1, \ \ \
C_{\bf rr'} \to C_2,
\label{C:def}
\end{equation}
in order to have $({\bf n_{rr'}},{\bf l_{rr'}})$ as canonical
variables of the resulting linear theory.
Here $Z$ is the coordination number, and sums
are taken over all nearest neighbors.
$C_1$ and $C_2$ do not depend on the bond ${\bf rr'}$ because
they must obey the translational symmetry of the paramagnetic phase, and
are determined self-consistently,
in a manner similar
to calculating the mass term in $1/N$ expansion of the quantum nonlinear
sigma model; they are not necessarily equal
to the equal time averages of the spin operators in Eqs.(\ref{C:def}).
The remaining fluctuation component
becomes the mode coupling term
written as $|{\bf n}|^2=1$ in the quantum nonlinear sigma model.
In our {\em linear} spin wave theory, the fluctuations of $C_{\bf rr'}$
around self-consistent averages $C_1$ and $C_2$
are ignored, which leads to the
following spectrum of noninteracting spin waves:
\begin{equation}
\epsilon_q = Z_\epsilon \sqrt{(1-\gamma_q)(1+\theta \gamma_q)},
\label{epsilonq}
\end{equation}
where $Z_\epsilon^2 = Z^2 C_1$,
$\theta = -C_2/C_1$, and
\begin{equation}
\gamma_q = \frac{1}{Z}\sum_\delta
\exp(i {\bf q} \delta).
\end{equation}
For the square lattice,
$\gamma_q = \frac{1}{2} (\cos(q_x)+\cos(q_y))$.

Our method of deriving the spin wave spectrum is similar in spirit to
the work of Villain \cite{Villain} and Haldane \cite{Haldane},
but uses an expressly isotropic set of variables.
Starykh recently pointed out an elegant
alternative method of derivation,
based on frequency moments of the dynamical susceptibility,
which will be described elsewhere \cite{Starykh}.
Villain \cite{Villain}, Young and Shastry \cite{Young:Shastry},
and others earlier obtained the same
$\epsilon_q$ as a function of wavevector,
but their $Z_\epsilon(T)$ and $\theta(T)$ are different from ours.

A similar form of $\epsilon_q$ arises in the
equations of motion closure methods (also called decoupling methods
\cite{Tyablikov})
that replace $A_{\bf rr'},B_{\bf rr'}$ in Eq.(\ref{motion})
by their equal time averages.
These methods are not equivalent to our theory, e.g. they
do not reproduce temperature dependences predicted by the
exact solution for $S=\infty$.

The dynamical spin susceptibility is derived using
the standard quantization procedure:
\begin{equation}
\chi({\bf q},i\omega_n) = -\frac{4}{3}\, U \times \frac{1-\gamma_q}
{\omega_n^2 + \epsilon_q^2},
\label{chiqw}
\end{equation}
where
\begin{equation}
U(T)=\frac{1}{2}\, \sum_{\delta} \langle {\bf S}_0 {\bf S}_{\delta}\rangle
\label{U:def}
\end{equation}
is the internal energy per spin.
$\chi({\bf q},i\omega_n)$ is subject to the
following two constraints:
\begin{equation}
T\sum_{n} \int \, \frac{d{\bf q}}{(2\pi)^d} \ \chi({\bf q},i\omega_n) =
\frac{1}{3}\, S(S+1),
\label{length:constraint}
\end{equation}
\begin{equation}
T\sum_{n} \int \, \frac{d{\bf q}}{(2\pi)^d} \ Z\gamma_q
\chi({\bf q},i\omega_n) = \frac{2}{3}U,
\label{U:constraint}
\end{equation}
where $\omega_n = 2\pi n T$ is the set of Matsubara frequencies,
and Eq.(\ref{U:constraint}) follows from Eq.(\ref{U:def}).

In the classical ($S=\infty$) limit,
Eqs.(\ref{epsilonq}-\ref{U:constraint})
reproduce the exact solution of the classical
Heisenberg chain by Fisher \cite{Fisher}; they remain exact
for Bethe lattices with arbitrary coordination number.

In two dimensions, Eqs.(\ref{epsilonq}-\ref{U:constraint})
depend on ${\bf q}$ only through $\gamma_q$, a prediction which
is not exact for lattices which contain loops,
such as the square lattice. We have verified that numerically calculated
observables for the Heisenberg models,
for example $\chi({\bf q})$ and $S({\bf q})$,
depend on ${\bf q}$ only through $\gamma_q$ to several per cent accuracy,
which is consistent with the small
probability, $1/32\approx 3\%$, for a four-step path on the square lattice
to form a loop.

{\bf Analytical Calculation of the Internal Energy}.
The closure of
Eqs.(\ref{epsilonq}-\ref{U:constraint}) requires one additional constraint
on the three variables $Z_\epsilon(T)$, $\theta(T)$, and $U(T)$.
In his exact solution, Fisher uses a special property of the
$S=\infty$ model, namely that the
internal energy $U(T)$ for a chain is the same as for an isolated spin pair.
Because this method does not apply for finite spin, we use a different
method based on the same assumption of linearity (no mode coupling) as
we used to derive $\epsilon_q$.

If a linear spin wave theory is a good approximation,
the internal energy at temperatures where occupation numbers are
sufficiently small must be given by (Anderson \cite{Anderson}):
\begin{equation}
U(T) = U_0 + N \int \,
\frac{d{\bf q}}{(2\pi)^d}
\frac{\epsilon^0_q}{\exp(\epsilon^0_q/T)-1},
\label{U}
\end{equation}
where the ground state energy $U_0$ is
\cite{Anderson,Oguchi,Igarashi}:
\begin{equation}
U_0 = -\frac{1}{2} Z S^2 \left(1+\frac{1-I_D}{S}+\frac{(1-I_D)^2}{4S^2} +
O\left(\frac{1}{S^3}\right)\right),
\label{U0}
\end{equation}
the $T=0$ spin wave spectrum:
\begin{equation}
\epsilon^0_q = ZS\left(1+\frac{1-I_D}{2S}  +
O\left(\frac{1}{S^2}\right) \right) \sqrt{1-\gamma_q^2},
\label{epsilon0}
\end{equation}
and
\begin{equation}
I_D = \int \, \frac{d{\bf q}}{(2\pi)^d} \sqrt{1-\gamma_q^2}
\approx  1-\frac{1}{2Z}.
\label{I_D}
\end{equation}

The asymptotic $T\to 0$ spin wave theory \cite{Anderson,Oguchi},
and hence the quantum nonlinear sigma model theory for
the renormalized classical regime \cite{CHN},
predict two gapless transverse spin wave modes per one
magnetic unit cell,
or per two spins, i.e. $N=1$ in Eq.(\ref{U}).

Our theory has no mode coupling, and therefore
does not distinguish between longitudinal and transverse modes
(such an approximation is valid for quantum spin models, but not for
classical models where $N=1$ should be used);
therefore, it must have three equivalent harmonic oscillator modes
per two spins, or $N=3/2$ per spin in Eq.(\ref{U}).
This can be seen, e.g.,  from the following argument: if each of the
spins is assigned to a dimer which it shares
with one of its neighbors, one set of canonical variables
$({\bf n_{rr'}},{\bf l_{rr'}})$ per each dimer correctly
counts the degrees of freedom,
three per dimer, or $N=3/2$ per unit cell.

\begin{figure}
\centerline{\epsfxsize=2.7in\epsfbox{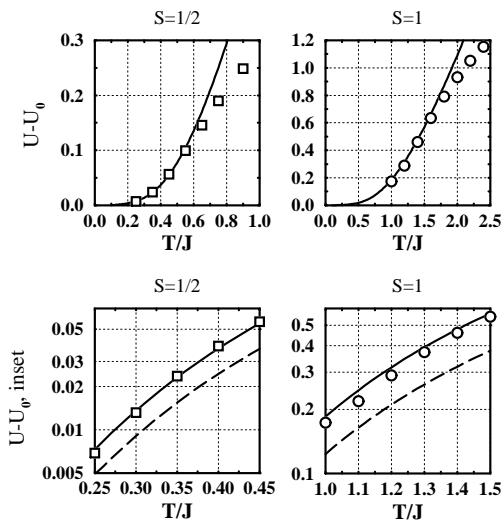}}
\caption{
Numerical results for the internal energy
measured with respect to the ground state $U(T)-U_0$
(squares, Ref.\protect\cite{Sandvik:Scalapino}; circles, this work)
are compared to Eqs.(\protect\ref{U}-\protect\ref{I_D})
without adjustable parameters.
Solid line corresponds to $N=3/2$ and dashed line to $N=1$
expected in the renormalized classical regime.
}
\label{fig:u}
\end{figure}

The crossover between $N=1$ and $N=3/2$ regimes is expected to occur
when the temperature is of order longitudinal mode gap.
Figure \ref{fig:u}
shows that our prediction of $N=3/2$
is consistent with the data down to the lowest temperatures
where numerical results are available, $T=0.25J$ for $S=1/2$ and
$T=J$ for $S=1$.

\begin{figure}
\centerline{
\epsfxsize=2.4in\epsfbox{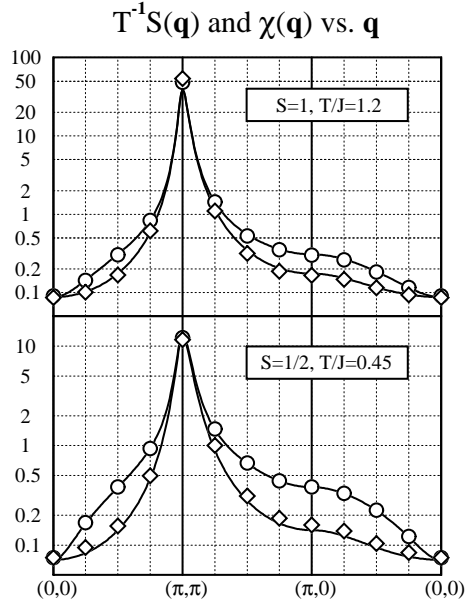}
}
\caption{
Numerical series expansions results for
the equal time correlator $T^{-1}S({\bf q})$ (circles) and the static
susceptibility $\chi({\bf q})$ (diamonds) across the Brillouin zone
are compared to our analytical theory
Eqs.(\protect\ref{chiq0},\protect\ref{Sq})
without adjustable parameters.
}
\label{fig:bz}
\end{figure}

\begin{figure}
\centerline{
\epsfxsize=2.9in\epsfbox{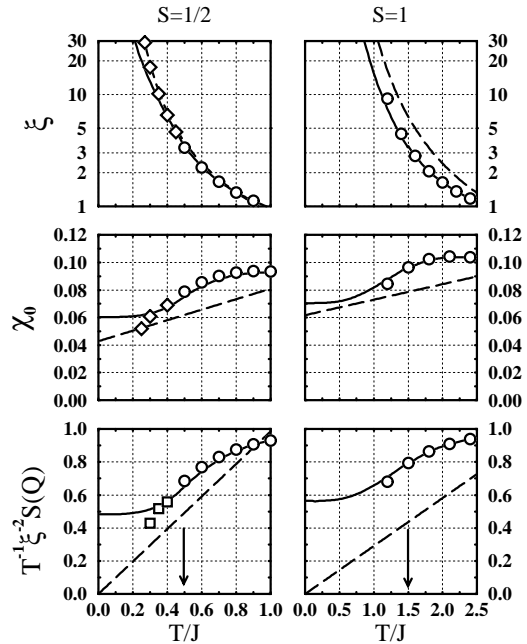}
}
\caption{
Numerical data for the correlation length $\xi$, bulk susceptibility
$\chi_0$, and normalized
Lorentzian amplitude $S_Q/(T\xi^2)$ for $S\!=\!1/2$
and $S\!=\!1$ models, from Ref.\protect\cite{Ding:Makivic} (diamonds),
Ref.\protect\cite{Sandvik:Scalapino} (squares), and this work (circles).
Solid lines is
the theory presented here; dashed lines is renormalized classical
theory \protect\cite{CHN} with numerical constants calculated in
\protect\cite{HN,CSY}. There are no adjustable
parameters; note however that our theoretical curves are completely
analytical below the temperature shown by arrow, and
use numerically calculated $U(T)$ as input above that temperature.
}
\label{fig:obs}
\end{figure}

\newpage

{\bf Comparison With the Numerical Data}.
The static susceptibility
\begin{equation}
\chi({\bf q},0) =
- \frac{4U}{3Z_\epsilon^2}
\times \frac{1}{1+\theta \gamma_q}
\label{chiq0}
\end{equation}
has exactly the same q-dependence as in the exact solution of the classical
Heisenberg chain, and is Lorentzian
near the ordering wavevector ${\bf Q}$. Accordingly,
the critical dimension of spin in this theory is equal to its
mean-field value $\eta=0$.
These predictions compare well with the actual nearly Lorentzian shape
of $\chi({\bf q},\omega\!=\!0)$ and a very small value of $\eta=0.023$
\cite{eta} known from numerical calculations.
In contrast with the static susceptibility,
the prediction for the equal-time correlator
\begin{equation}
S({\bf q}) = - \frac{2}{3}\, U \times \frac{1-\gamma_q}
{\epsilon_q \tanh(\epsilon_q/2T)}
\label{Sq}
\end{equation}
does not have Lorentzian shape near ${\bf Q}$.
These analytically calculated $\chi({\bf q})$ and $S({\bf q})$ are
plotted in Fig.\ref{fig:bz} as a function of wavevector for the lowest
temperature where
numerical results for q-dependence are available for comparison.
We find good agreement
of analytical and numerical curves, achieved without
adjustable parameters.

Analytical and numerical results for temperature dependences of the
correlation length $\xi(T) = \frac{1}{2} \sqrt{\theta/(1-\theta)}$
(defined from the long-distance decay of correlations
$\sim \exp(-R/\xi)$),
bulk susceptibility $\chi_0(T)$, and normalized
Lorentzian amplitude $S_Q/(T\xi^2)$
are plotted in Fig.\ref{fig:obs}.
Good agreement between numerical
and analytical results is observed in most of the temperature range,
except for the lowest temperatures for $S=1/2$
(roughly $\xi>10$) where all three quantities begin to deviate from the
theoretical predictions. The deviations are most likely
caused by a crossover to the renormalized classical \cite{CHN}
behavior. Evidently, the crossover is completed at still lower temperatures,
where no numerical data is currently available; analysis
of the experiment remains the only currently available
alternative for studying this crossover further.

{\bf Summary}.
We have developed an isotropic  spin wave
theory of short range magnetic
order and compared it with the independently obtained
numerical data for $S\!=\!1/2$ and $S\!=\!1$
square lattice antiferromagnets.
Unlike the quantum nonlinear sigma or any other continuous model,
our theory does not require the correlation length to be much
longer than the lattice spacing. Furthermore, because
our theory is valid for all wavevectors throughout the Brillouin zone,
it is suitable for calculating
those quantities that are not dominated by correlations near the
ordering wavevector. Its primary drawback, compared to the
quantum nonlinear sigma model, is linearity
and the resulting lack of mode coupling or dissipation
(the continuous limit of our theory
is the $O(\infty)$ quantum nonlinear sigma
model, and not the physical $O(3)$ model).

We found that the
analytical theory agrees with the numerical data without adjustable
parameters. In most cases, the agreement is within few
per cent in the range of applicability, which for these two models
approximately
corresponds to the correlation length of less than ten lattice
spacings. We expect this theory to be applicable for a variety of
other magnets,
and are planning to pursue further studies in this direction.

{\bf Acknowledgements}. We thank R.J. Birgeneau, A.V. Chubukov, M. Greven,
T. Jolicoeur, S. Sachdev, S. Sondhi, and O. Starykh
for valuable discussions,
and A. Sandvik for providing numerical data for comparisons.
A.S. is an A.P. Sloan Research Fellow.
This work is supported by NSF under Grant No. DMR93-18537, and
in part under Grant No. PHY94-07194 through the Institute for Theoretical
Physics at UC Santa Barbara.
A.S. is grateful to T. Jolicoeur for hospitality during his stay at
CEA-Saclay, France, where part of this work was done.


\end{document}